\documentclass[conference]{IEEEtran}
\usepackage[T1]{fontenc}
\usepackage{cite}
\usepackage{url}
\usepackage{amsmath,amssymb,amsfonts}
\usepackage{algorithmic}
\usepackage{graphicx}
\graphicspath{{./images/}}
\usepackage{textcomp}
\usepackage{xcolor}
\usepackage{microtype}
\usepackage{multirow}
\usepackage{booktabs}   
\usepackage{adjustbox}
\def\BibTeX{{\rm B\kern-.05em{\sc i\kern-.025em b}\kern-.08em
    T\kern-.1667em\lower.7ex\hbox{E}\kern-.125emX}}

\IEEEoverridecommandlockouts

\begin{document}

\title{DualAttWaveNet: Multiscale Attention Networks for Satellite Interference Detection}

\author{
    \IEEEauthorblockN{
        Chunyu Yang,
        Boyu Yang,
        Kun Qiu,
        Zhe Chen,
        Yue Gao
    }
    \IEEEauthorblockA{
        School of Computer Science, Fudan University, China\\
        \{22307140114, 24110240144\}@m.fudan.edu.cn, \{qkun, zhechen, gao.yue\}@fudan.edu.cn
    }
    \thanks{This work was supported by the Fudan Undergraduate Research Opportunities Program (FDUROP) under Grant No.\,24198.}
}

\maketitle

\begin{abstract}
    The escalating overlap between non-geostationary orbit (NGSO) and geostationary orbit (GSO) satellite frequency allocations necessitates accurate interference detection methods that address two pivotal technical gaps: computationally efficient signal analysis for real-time operation, and robust anomaly discrimination under varying interference patterns.  Existing deep learning approaches employ encoder-decoder anomaly detectors that threshold input-output discrepancies for robustness. While the transformer-based TrID model achieves state-of-the-art performance (AUC: 0.8318, F1: 0.8321), its multi-head attention incurs prohibitive computation time, and its decoupled training of time-frequency models overlooks cross-domain dependencies. To overcome these problems, we propose DualAttWaveNet. A bidirectional attention fusion layer dynamically correlates time-domain samples using parameter-efficient cross-attention routing. A wavelet-regularized reconstruction loss enforces multi-scale consistency.  We train the model on public dataset which consists of 48 hours of satellite signals. Experiments show that compared to TrID, DualAttWaveNet improves AUC by 12\% and reduces inference time by 50\% to 540ms per batch while maintaining F1-score.

\end{abstract}


\section{Introduction}
\label{sec:intro}

The rapid increase of low Earth orbit (LEO) satellite systems presents significant challenges for next-generation communication networks. Industry projections indicate over 20,000 satellites will be deployed by leading operators, including SpaceX's Starlink \cite{starlink} and Starshield \cite{spacex_starshield}, as well as Eutelsat OneWeb \cite{oneweb}, by the end of the decade. These mega-constellations have become critical infrastructure for global connectivity, simultaneously commercializing space-based communications and expanding broadband access to previously underserved regions \cite{reddyLowEarthOrbit2023}. However, this exponential growth in satellite deployment creates fundamental technical obstacles—particularly the rising risk of spectrum overlap between LEO and geosynchronous orbit (GSO) satellites—necessitating scalable interference management frameworks that can evolve alongside expanding LEO networks.

Current research in satellite interference management generally follows three approaches. Preventive measures aim to reduce risks before system deployment \cite{sharmaInlineInterferenceMitigation2016, liOptimalBeamPower2019}, static mitigation methods are applied after interference occurs \cite{wangCoFrequencyInterferenceAnalysis2020, zhangSpectralCoexistenceLEO2018}, and simulation-based models predict interference using limited time or location samples. While useful in controlled settings, these methods face significant limitations in real-world environments \cite{yunDynamicDownlinkInterference2023}. Preventive measures rely on assumptions that often prove invalid after deployment, static mitigation strategies cannot adapt to dynamic interference patterns, and simulation models frequently fail to account for unexpected environmental factors such as solar radiation fluctuations or atmospheric variations \cite{facskoSpaceWeatherEffects2023}. Furthermore, traditional detection techniques that depend on fixed thresholds or static signal features cannot reliably identify complex, real-time interference. Addressing these shortcomings requires new detection solutions that achieve both rapid real-time response and consistent accuracy across diverse interference conditions.

Approaches to interference detection in satellite communications can be broadly categorized into traditional analytical methods and machine learning (ML)-based techniques. Conventional methods primarily employ energy detection (ED), which calculates signal energy over predefined intervals for threshold-based anomaly identification \cite{kay2009fundamentals}, or exploit spectral cyclostationary features to distinguish interference from periodic signals \cite{experimentalCyclostationary}. However, these approaches require manually calibrated thresholds and demonstrate limited sensitivity in low signal-to-noise ratio (SNR) environments. In contrast, ML-driven methods eliminate such constraints by automatically learning discriminative interference signatures from raw data. Classification-based solutions utilize deep neural networks to establish decision boundaries \cite{pellacoSpectrumPredictionInterference2019}, while encoder-decoder architectures treat detection as an anomaly identification task by reconstructing interference-free waveforms from corrupted inputs. Recent advancements have further deployed transformer models to capture long-range spectral dependencies, improving detection accuracy for persistent anomalies \cite{saifaldawlaGenAIBasedModelsNGSO2024}.

Despite these advancements, several critical challenges persist in existing interference detection methods. First, threshold-dependent traditional approaches—exemplified by energy detection—show degraded reliability in low-SNR regimes, where static thresholds fail to dynamically adapt to noise fluctuations or interference intensity variations, resulting in frequent false negatives under rapidly evolving orbital conditions \cite{saifaldawlaGenAIBasedModelsNGSO2024}. Second, while attention-driven architectures achieve state-of-the-art detection sensitivity, their computational overhead from multi-head attention mechanisms creates significant latency during both training and inference, making them unsuitable for resource-constrained satellite edge devices. Third, contemporary deep learning models often process time-domain and frequency-domain signal representations in isolation by training separate networks for each modality. This decoupled approach yields suboptimal performance—time-domain models achieve an AUC of only 0.8318 while frequency-domain models score a mere 0.7106, fundamentally limiting detection capability.

\begin{figure}[t!]
    \centering
    \includegraphics[width=\linewidth]{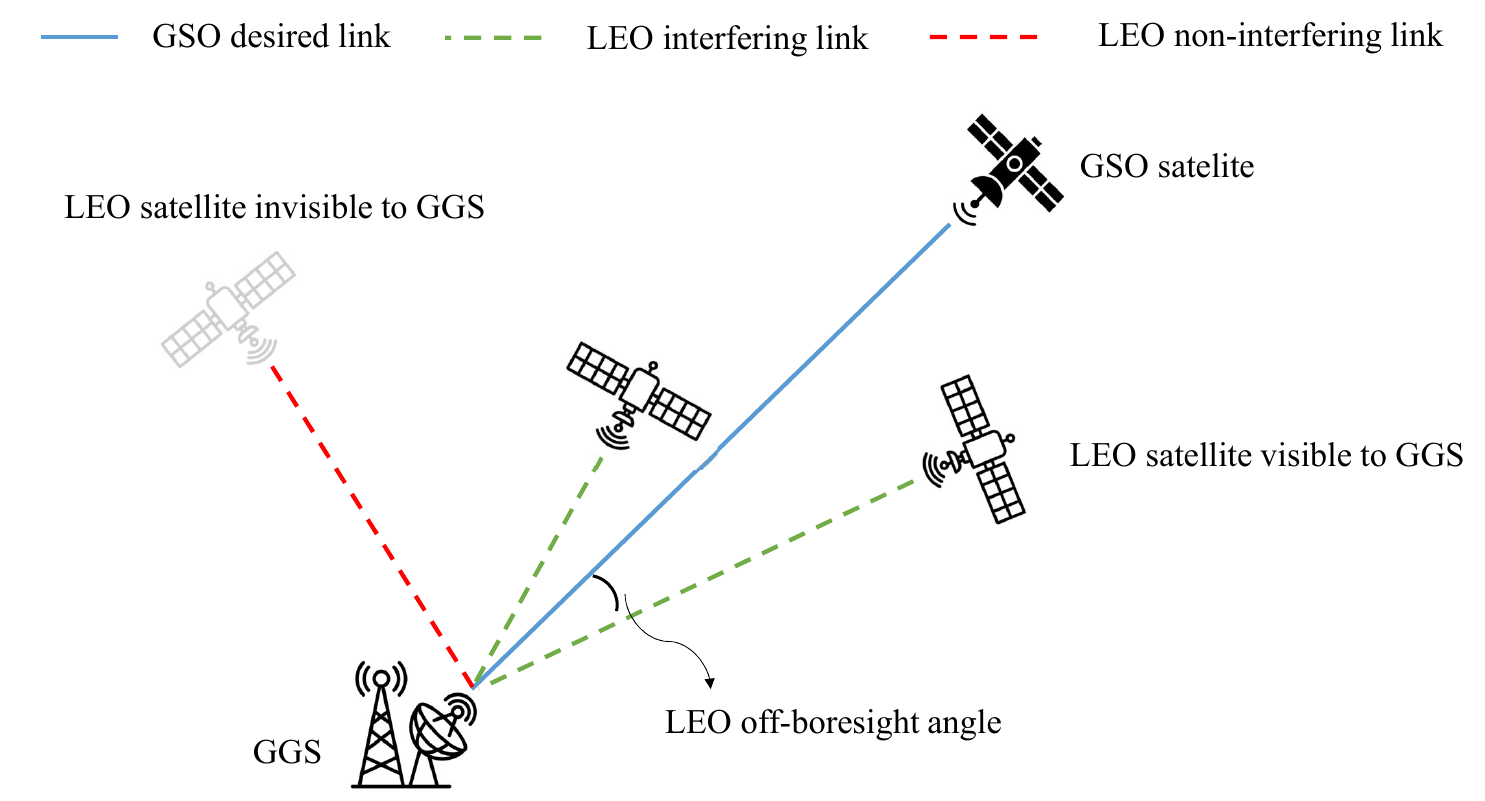}
    \caption{Interference scenario between GSO and LEO satellite systems.}
    \label{fig:interference-scenario}
\end{figure}

To address these challenges, we propose \textbf{DualAttWaveNet}, a unified model that integrates cross-domain signal fusion for real-time interference detection. Our primary contributions are threefold:

\begin{enumerate}
    \item The proposed architecture jointly processes time-domain samples and their frequency-domain representations through a bidirectional attention mechanism. This design eliminates the need for explicit multi-head computation while enabling adaptive correlation learning between spectral and temporal features.
    \item To enhance robustness against minor signal fluctuations, we introduce a wavelet-constrained reconstruction loss. This is achieved by applying discrete wavelet transform (DWT) decomposition to both raw and reconstructed signals, enforcing cross-band consistency through multi-scale subspace constraints.
    \item Comprehensive evaluation on a public satellite communication dataset (48 hours duration) demonstrates DualAttWaveNet's superiority: 12\% higher AUC and 50\% faster inference compared to state-of-the-art baselines, while maintaining competitive F1-score.
\end{enumerate}

\begin{figure*}[tb]
    \centering
    \includegraphics[width=0.9\linewidth]{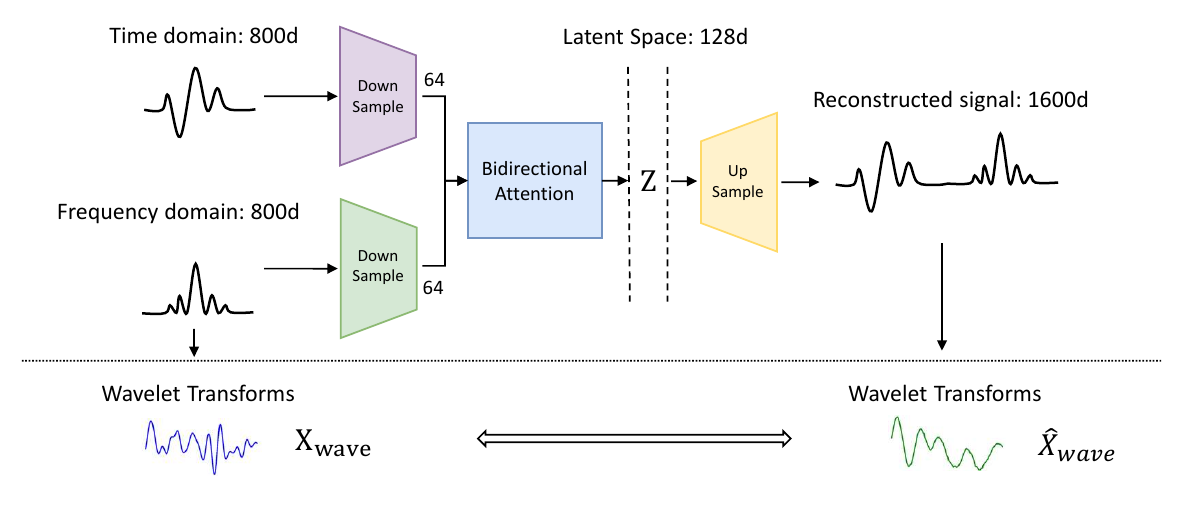}
    \caption{Summary of our approach. Architecture of the cross-modal signal enhancement model showing dual-branch processing: time domain (800d) and frequency domain (800d) inputs are downsampled to 64d, fused through bidirectional attention into a 128d latent space ($\mathcal{Z}$), then upsampled to produce a reconstructed 1600d signal. Wavelet transforms ($X_\text{wave}$ and $\hat{X}_{\text{wave}}$) are applied to the original and reconstructed signals.}

    \label{fig:overview}
\end{figure*}

The remainder of this paper is organized as follows. Section 2 provides background information and motivations. Section 3 details the architecture of DualAttWaveNet. Section 4 presents our experimental results and analysis. Finally, Section 5 concludes the paper with a summary of our findings and directions for future work.

\section{Background}
\label{sec:background}

\subsection{Interference Scenario}

Satellite communication systems using the same frequency band often interfere with each other. As shown in Fig.~\ref{fig:interference-scenario}, we focus on two types of satellites. GSO Satellite flies in a fixed position 36,000 km above Earth. It serves as the main signal source for a ground station (GGS). LEO Satellites move rapidly at 500-2,000 km altitudes. Their signals can interfere with GSO signals due to spectrum overlap. The composite received waveform of GGS contains both desired carrier signals and interference.

The received carrier power $C$ from a geosynchronous orbit (GSO) satellite is calculated based on classical satellite link budget principles:
\begin{equation}
    C = \frac{\text{EIRP}_{\text{gso}} \cdot G_{\text{r, gso}}(\theta_0)}{L_{\text{FS, gso}} \cdot L_{\text{add}}}
    \label{eq:carrier_power}
\end{equation}
where $\text{EIRP}$ is GSO satellite equivalent isotropic radiated power, $G_{\text{r, gso}}(\theta_0)$ denotes the maximum receive antenna gain at boresight angle $\theta_0$, $L_{\text{FS, gso}}$ represents free-space path loss, and $L_{\text{add}}$ accounts for aggregate atmospheric impairments.

Interference from $K$ low Earth orbit (LEO) satellites is modeled as the superposition of individual interference terms:
\begin{equation}
    I_k = \frac{\text{EIRP}_k \cdot G_{\text{r, k}}(\theta_k) \cdot B_{\text{adj, k}}}{L_{\text{FS, k}} \cdot L_{\text{add}}}
    \label{eq:interference_power}
\end{equation}

The angular gain term $G_{\text{r, k}}(\theta_k)$ reflects spatial relationships caused by LEO orbital motion, while $B_{\text{adj, k}} \in [0,1]$ is the spectral overlap between GSO and LEO transmissions.

The signal received by physical layer at the GGS has three components:
\begin{align}
    y(t) =\, & x(t)\sqrt{\text{CNR}}\, (\text{Desired GSO}) \nonumber                                                 \\[0.5em]
             & + \sum_{k=1}^{K} I_k(t)e^{j2\pi \Delta f_k t}\sqrt{\text{INR}_k}\, (\text{LEO interference}) \nonumber \\[0.5em]
             & + \zeta(t)\, (\text{Thermal noise})
\end{align}
where $x(t)$  is the desired GSO signal, $\Delta f_k = f_{\text{c},k} - f_{\text{c,gso}}$ captures carrier frequency offsets from Doppler effects. The exponential terms induce time-varying phase rotations proportional to relative satellite motion. Here, $\text{CNR}$ (Carrier-to-Noise Ratio) and $\text{INR}_k$ (Interference-to-Noise Ratio) respectively characterize the desired signal quality and interference intensity relative to the noise floor.

Dual signal representations are derived for machine learning processing:
\begin{itemize}
    \item Time-domain: $y^A$ captures instantaneous amplitude variations through uniform sampling
    \item Frequency-domain: Welch's power spectral density estimation generates logarithmic magnitude spectra via overlapping windowed transforms: $y^F = 10\log_{10}(\phi(y(t)))$
\end{itemize}

\subsection{Deep Learning Approaches for Interference Detection}
\label{ssec:related_works}

Recent advances in machine learning have significantly shifted the landscape of satellite interference detection, moving away from traditional signal processing techniques toward anomaly-based approaches. One widely adopted method uses encoder-decoder frameworks that are trained to reconstruct clean, interference-free waveforms from raw signal inputs. The core idea is that interference introduces deviations from the expected signal, which can be identified by comparing input and output waveforms.

Unlike regression-based models that struggle to capture the diversity of interference types, reconstruction-based methods are more flexible, as they directly highlight anomalies through mismatches between original and reconstructed signals. Early work by Pellaco et al. \cite{pellacoSpectrumPredictionInterference2019} introduced an LSTM autoencoder (LSTMAE) for identifying interference in non-geostationary satellite data. However, the sequential nature of LSTM limits parallelism, making the model less suitable for real-time systems. To address this, Saifaldawla et al. \cite{saifaldawlaConvolutionalAutoencodersNonGeostationary2024} proposed convolutional autoencoders (CAE) that process time domain signal directly.

More recently, transformer-based architectures have been adopted to capture long-range dependencies in signal data \cite{saifaldawlaGenAIBasedModelsNGSO2024}. These models leverage self-attention mechanisms to identify global interference patterns that traditional convolutional or recurrent layers may miss. A typical setup encodes each input—consisting of 800 sequential time-domain samples—into a 64-dimensional latent vector using stacked transformer layers. The decoder reconstructs the waveform through learned upsampling operations. During training, a mean squared error (MSE) loss between the original and reconstructed signal guides the model to learn interference-free patterns. At inference time, samples containing interference produce larger reconstruction errors, which are flagged as anomalies when they exceed a chosen threshold.

Despite their performance, transformer-based models come with drawbacks. Their computational complexity scales quadratically with input length, making them difficult to apply in wideband satellite systems. Moreover, many current approaches treat time and frequency information separately, missing out on the potential benefits of jointly modeling these domains. This limitation reduces the ability to learn cross-domain features that could improve robustness under diverse interference conditions.

\section{Overall Design}
\label{sec:model}

\subsection{Proposed Deep Learning Model}
\label{subsec:proposed_model}

In light of the challenges discussed in Section \ref{sec:background}, specifically the prohibitive computational cost of multi-head attention and the need for higher classification accuracy—we propose DualAttWaveNet. It is an autoencoder that accepts dual-domain inputs and reconstructs both representations concurrently.

As illustrated in \figurename~\ref{fig:overview}, the architecture features separate encoders and decoders for the time-domain and frequency-domain signals. A key innovation in our design is the incorporation of a bidirectional attention fusion module, which replaces conventional multi-head attention. This module leverages single-head dot-product attention with spatial reduction, thereby minimizing computational overhead while efficiently capturing cross-domain dependencies. The fusion of features prior to decoding is critical for achieving superior reconstruction accuracy, which in turn improves the overall detection performance.

The input for both domains is presented as tensors of shape $B \times 800$, with $B$ denoting the batch size. Following dedicated convolution modules that serve as downsampling layers, each input is transformed into a $B \times 64$ latent representation. These latent features are then processed by the bidirectional attention module, where cross-modal interactions are computed in both directions. The resultant features are concatenated and fed into upsampling layers that reconstruct the outputs, yielding a final tensor of shape $B \times 1600$.

Furthermore, to directly address issues related to classification accuracy, the model is trained using a composite loss function that integrates the conventional mean squared error (MSE) with a novel wavelet-domain regularization term, dicussed in Section~\ref{subsec:wavelet}. This wavelet loss enforces multi-scale consistency during reconstruction, ensuring that both temporal features and spectral distributions are preserved.

\subsection{Bidirectional Attention}
\label{subsec:bi_attn}

\begin{figure}[tb]
    \centering
    \includegraphics[width=0.65\linewidth]{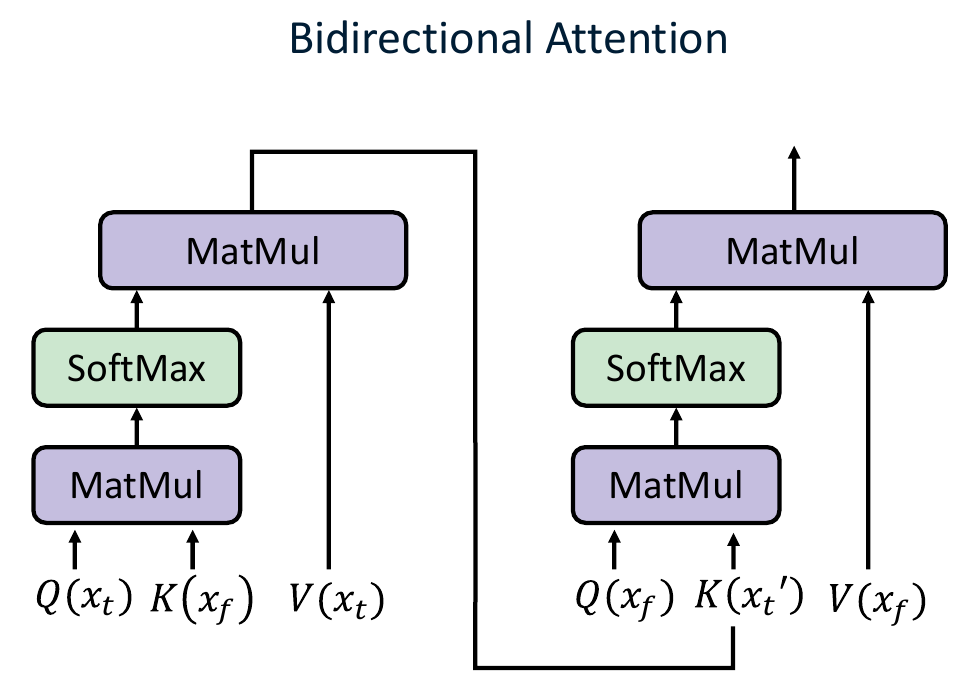}
    \caption{Bidirectional attention mechanism in DualAttWaveNet. Both attention phases share the same parameters $Q$, $K$, and $V$, implemented as 1D convolutions to reduce channel dimensions.}
    \label{fig:bidirectional-attention}
\end{figure}

We introduce a parameter-efficient mutual attention mechanism for cross-modal feature fusion. Unlike traditional multi-head attention \cite{vaswaniAttentionAllYou2017}, our method uses single-head dot-product attention with spatial reduction to lower computational costs while preserving inter-domain alignment, as shown in \figurename~\ref{fig:bidirectional-attention}.

For input feature maps $\mathbf{X} \in \mathbb{R}^{B \times C \times L}$ (time domain) and $\mathbf{Y} \in \mathbb{R}^{B \times C \times L}$ (frequency domain), the mutual attention operator is defined as:
\begin{equation}
    \begin{aligned}
        \text{MutualAttn}(\mathbf{X}, \mathbf{Y})    & = \mathbf{X} + \gamma \cdot \text{AttentionGate}(\mathbf{X}, \mathbf{Y})                         \\
        \text{AttentionGate}(\mathbf{X}, \mathbf{Y}) & = \mathbf{V}_y \cdot \text{Softmax}\Bigl(\frac{\mathbf{Q}_x\, \mathbf{K}_y^\top}{\sqrt{d}}\Bigr)
    \end{aligned}
\end{equation}
where $\gamma$ is a learnable scalar (initialized to 0) and $\mathbf{Q}_x = \mathcal{W}_Q(\mathbf{X}) \in \mathbb{R}^{B \times L \times \frac{C}{8}}$, $\mathbf{K}_y = \mathcal{W}_K(\mathbf{Y}) \in \mathbb{R}^{B \times \frac{C}{8} \times L}$, and $\mathbf{V}_y = \mathcal{W}_V(\mathbf{Y}) \in \mathbb{R}^{B \times C \times L}$. Here, $\mathcal{W}_{Q,K,V}$ are 1D convolutional layers with kernel size 1 that reduce channel dimensions by a factor of 8.

The reduced representations are multiplied to obtain an $L \times L$ affinity matrix, capturing position-wise correlations between the two modalities. A row-wise softmax is then applied to normalize the scores.

To ensure training stability, the residual connection is initially dampened ($\gamma=0$) and gradually increased. The same attention mechanism is applied symmetrically in both directions:
\begin{equation}
    \begin{aligned}
        \widehat{\mathbf{X}} & = \text{MutualAttn}(\mathbf{X}, \mathbf{Y})\,,           \\
        \widehat{\mathbf{Y}} & = \text{MutualAttn}(\mathbf{Y}, \widehat{\mathbf{X}})\,.
    \end{aligned}
\end{equation}

This bidirectional design, with shared parameters, efficiently refines both modalities without significant computational overhead.

\subsection{Wavelet-Domain Spectral Regularization}
\label{subsec:wavelet}

Standard MSE loss alone does not fully capture a signal's complex structure. To address this, we introduce a wavelet loss that enforces reconstruction consistency across multiple scales. For an input tensor $\mathbf{x}\in\mathbb{R}^{B\times C\times L}$ (with batch size $B$, channels $C$, and length $L$), we build a learnable filter bank $\mathcal{F}\in\mathbb{R}^{S\times 1\times K}$ parameterized by
\begin{equation}
    \label{eq:wavelet}
    \mathcal{W}_s = \operatorname{Norm}\Bigl(\cos\Bigl(\frac{t}{s}\tau\Bigr) \odot \mathcal{G}(\tau,s)\Bigr),
\end{equation}
where $\mathcal{G}(\tau,s)=\exp\left(-\frac{\tau^2}{2s^2}\right)$, $s\in\mathbb{S}$ are the scale parameters, $K=4s_{\text{max}}$ sets the kernel size, and $\tau$ is the index within $-2K$ and $2K$. Each filter is $L_2$-normalized to maintain energy consistency.

\begin{figure}[tb]
    \centering
    \includegraphics[width=0.9\linewidth]{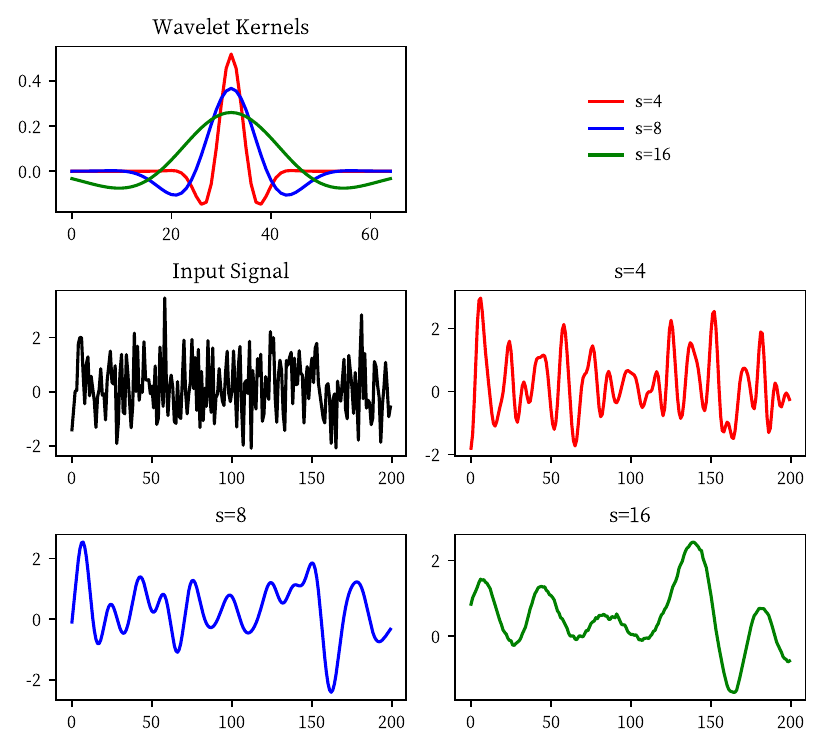}
    \caption{(Top) Normalized Morlet wavelet kernels spanning scales \( s \in \{4, 8, 16\} \), designed to capture variable temporal-spectral features. (Bottom) Example analysis of a synthetic signal: Gaussian white noise (top-left) decomposed into distinct frequency components (red: \( s=4 \), high-frequency; blue: \( s=8 \), mid-frequency; green: \( s=16 \), low-frequency). Larger scales broaden temporal support for low-frequency emphasis, while smaller scales localize high-frequency transients.
    }

    \label{fig:wavelet-transform}
\end{figure}

We implement the discrete wavelet transform as an asymmetric depth-wise 1D convolution:
\begin{equation}
    \begin{aligned}
        \mathbf{X}_{\text{wave}} & = \text{Conv1D}(\mathbf{X}, \mathcal{F})                                                            \\
                                 & = \bigcup_{s\in\mathbb{S}} \mathbf{X}\ast \mathcal{W}_s \in \mathbb{R}^{B\times C\times S\times L}.
    \end{aligned}
\end{equation}
Reflection padding is used to mitigate boundary artifacts while preserving temporal resolution. \figurename~\ref{fig:wavelet-transform} illustrates example wavelet kernels and their filtering effects.

Our overall loss function combines standard MSE with the wavelet loss:
\begin{equation}
    \mathcal{L} = \lambda_1\|\mathbf{\hat{X}} - \mathbf{X}\|_2^2 + \lambda_2\sum_{s\in\mathbb{S}} \|\mathbf{\hat{X}}_{\text{wave}}^{(s)} - \mathbf{X}_{\text{wave}}^{(s)}\|_2^2,
\end{equation}
where $\lambda_1$ and $\lambda_2$ are empirically tuned weights. This dual-domain regularization ensures that both temporal features and spectral content are accurately preserved in the reconstruction.

\subsection{Threshold Selection}

The final output of the model is a reconstructed signal. To determine whether the input signal contains interference, we compare the reconstructed signal with the original signal. If the difference is larger than a predefined threshold, we classify the input signal as interference. The threshold is determined by the following formula:

\begin{equation}
    \Gamma_{\text{th}} = \mu + \text{std}(\mathbf{L})
\end{equation}
where $\mu$ is the mean of the loss in validation set.  $\text{std}(\mathbf{L})$ is the standard deviation of the loss. This strategy assumes that the loss of the model on the validation set is normally distributed, and the threshold is set to be one standard deviation above the mean. Any loss above this threshold is considered outliers, hence containing interference.

\section{Experiments}
\label{sec:experiments}

% \subsection{Data Configuration}

% The synthetic dataset is generated through a 48-hour MATLAB simulation sampling Ku-band (10.7-12.7 GHz) interference scenarios at 10-second intervals, producing 17,281 temporal snapshots following \cite{saifaldawlaGenAIBasedModelsNGSO2024}. Each instance contains synchronized time-domain and frequency-domain representations: an 800-point waveform captures signal amplitudes, while an 800-bin spectral magnitude is derived via FFT processing.

% Binary classification labels are assigned through link budget analysis, where class 0 denotes non-interference scenarios ($\text{INR} < \Gamma_{\text{th}}$) below the system protection threshold, and class 1 indicates substantial interference ($\text{INR} \geq \Gamma_{\text{th}}$) exceeding operational limits. We normalize the input signals in both domains separately to zero mean and unit variance.

% The dataset is partitioned under anomaly detection constraints, with training (11,509 samples) and validation (1,302 samples) sets containing exclusively non-interference data (class 0). The test set comprises balanced proportions of 2,235 class 0 and 2,235 class 1 instances. The simulation incorporates time-varying link losses with 0-9 dB range, extreme interference cases reaching peak aggregate INR of 32.47 dB, with background CNR fluctuations between 6.40-15.40 dB.

\subsection{Data Configuration}

The synthetic dataset was generated via a 48-hour MATLAB simulation, sampling Ku-band (10.7–12.7 GHz) interference scenarios every 10 seconds, producing 17,281 temporal snapshots as described in \cite{saifaldawlaGenAIBasedModelsNGSO2024}. Each instance includes synchronized time-domain and frequency-domain representations: an 800-point waveform captures signal amplitudes, and an 800-bin spectral magnitude is obtained through FFT processing.

Binary classification labels are determined by link budget analysis, where class 0 represents non-interference scenarios ($\text{INR} < \Gamma_{\text{th}}$) and class 1 denotes substantial interference ($\text{INR} \geq \Gamma_{\text{th}}$). Input signals in both domains are normalized separately to zero mean and unit variance.

The dataset is partitioned with anomaly detection in mind. The training (11,509 samples) and validation (1,302 samples) sets consist exclusively of non-interference data (class 0), while the test set is balanced with 2,235 samples each of class 0 and class 1. The simulation incorporates time-varying link losses (0–9 dB), extreme interference cases with peak aggregate INR of 32.47 dB, and background CNR fluctuations between 6.40 and 15.40 dB. All experiments were conducted on a laptop equipped with an RTX 3050Ti GPU (4GB VRAM).

% \subsection{Baseline Models}

% We use the following models as baselines to benchmark our framework: \textbf{LinearA}E with fully-connected encoder-decoder architectures as a simplistic reconstruction baseline; \textbf{CNNAE} utilizing 1D convolutional layers for encoder and MLP for decoder; \textbf{CNNAE+Attention} augmenting CNNAE with temporal self-attention modules; domain-specific \textbf{TrID} embedding spectral correlation priors; and \textbf{Transformer AE} constructed with stacked multi-head attention layers for global context modeling. All baselines adhere to the standard autoencoder paradigm that reconstructs inputs from bottleneck embeddings, implemented with identical training protocols and hyperparameter tuning strategies for fair comparison.

\subsection{Baseline Models}

We compare our method against several state-of-the-art reconstruction-based models: \textbf{CNNAE} utilizing 1D convolutional layers for the encoder and an MLP for the decoder; \textbf{CNNAE+Attention} that augments CNNAE with temporal self-attention modules; domain-specific \textbf{TrID} embedding spectral correlation priors; and the \textbf{Transformer AE} constructed with stacked multi-head attention layers for global context modeling. All baselines follow the standard autoencoder paradigm where inputs are reconstructed from bottleneck embeddings, and each model is trained using identical protocols and hyperparameter tuning strategies for fair comparison.

\subsection{Evaluation Results}

As shown in Table \ref{tab:main_results}, DualAttWaveNet achieves state-of-the-art performance across all key metrics. It records the highest AUC score (0.9327), surpassing the best baseline by 1.6\% and showing a 36.9\% absolute improvement over the Transformer AE (0.6812). Moreover, our framework completes inference in 0.5409s—46\% of TrID's 1.0156s—while maintaining competitive F1 score parity (0.8351) with lightweight models like CNNAE (0.8054). Notably, the F1 score of DualAttWaveNet exhibits only a minor 0.3\% performance drop relative to the task-specialized TrID (0.8321), effectively balancing precision and recall through integrated dual-attention mechanisms.

Baseline analyses reveal key architectural limitations: the Transformer AE's stacked multi-head attention layers incur substantial latency (1.8354s) and suffer from overparameterization, resulting in significant performance degradation (–22.5\% F1 compared to our method). Although domain-specific TrID achieves near-parity accuracy (0.8318), its AUC score of 0.8318 indicates an overfitting to spectral correlation priors. Similarly, the temporal attention in CNNAE+Attention increases computation time by 3.6× over vanilla CNNAE (0.5969s vs. 0.1654s) while degrading AUC by 5.2\%, highlighting the inefficiency of ill-configured attention modules compared to our optimized dual-attention design.

\begin{figure}[t]
    \centering
    \includegraphics[width=0.7\linewidth]{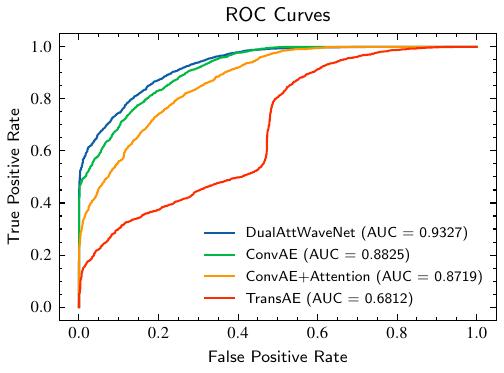}
    \caption{ROC curves for DualAttWaveNet and baseline models. DualAttWaveNet achieves the highest AUC score.}
    \label{fig:roc_comparison}
\end{figure}

\begin{figure*}[tb]
    \centering
    \includegraphics[width=\linewidth]{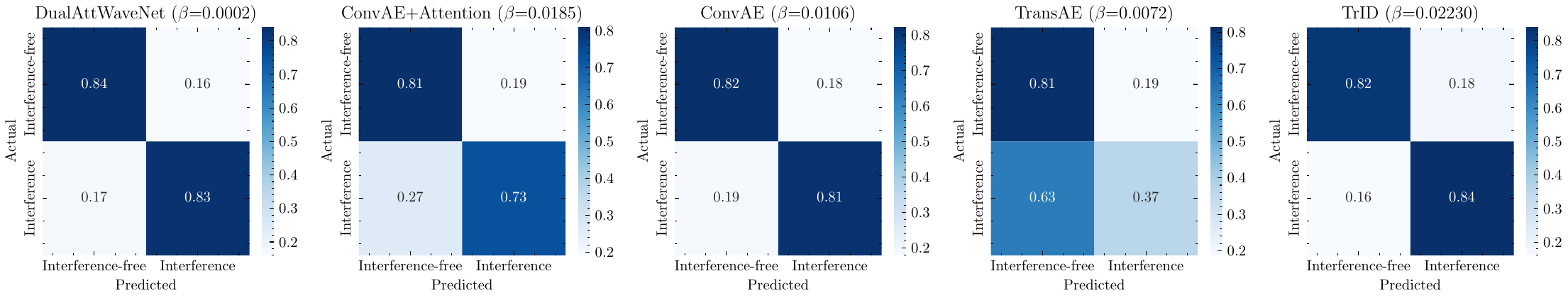}
    \caption{Confusion matrix for DualAttWaveNet and other baseline models.}
    \label{fig:confusion_matrix}
\end{figure*}

As visualized in \figurename~\ref{fig:roc_comparison}, the ROC curves clearly illustrate the discriminative power among competing models. DualAttWaveNet's left-skewed curve (AUC = 0.9327) dominates the upper-left quadrant, achieving an 83.5\% true positive rate at less than 10\% false positives. In contrast, the ROC trajectory of CNNAE+Attention drifts rightward (AUC = 0.8719), and Transformer AE hovers near the chance line (AUC = 0.6812), reflecting their inability to learn robust decision boundaries despite higher computational cost. \figurename~\ref{fig:confusion_matrix} further reveals class-specific recognition patterns across the evaluated models.

\begin{table}[t]
    \caption{Performance Comparison of DualAttWaveNet Against Baseline Models}
    \label{tab:main_results}
    \centering
    \resizebox{\linewidth}{!}{
        \begin{tabular}{lcccc}
            \toprule
            \textbf{Model}  & \textbf{Accuracy (\%) } $\uparrow$ & \textbf{F1 Score} $\uparrow$ & \textbf{AUC}$\uparrow$ & \textbf{Time(s)} $\downarrow$ \\
            \midrule
            DualAttWaveNet  & 0.8351                             & 0.8351                       & 0.9327                 & 0.5409                        \\
            \cmidrule{1-5}
            % LinearAE        & 0.8149                             & 0.8149                       & 0.9176                 & 0.0966                        \\
            CNNAE           & 0.8020                             & 0.8054                       & 0.8825                 & 0.1654                        \\
            CNNAE+Attention & 0.7695                             & 0.7691                       & 0.8719                 & 0.5969                        \\
            TrID            & 0.8318                             & 0.8321                       & 0.8318                 & 1.0156                        \\
            Transformer AE  & 0.6812                             & 0.5921                       & 0.6812                 & 1.8354                        \\
            \bottomrule
        \end{tabular}}
\end{table}

% The temporal localization parameter $t$ in Eq. \ref{eq:wavelet} governs the precision of interference feature extraction. Our experiments show that setting $t=2$ yields peak performance (0.9327 AUC). Values of $t=1$ cause a 0.7\% AUC drop due to over-localization, while larger values (e.g., $t=16$) blur critical temporal details, leading to a 4.5\% AUC degradation. These findings confirm that $t=2$ is the optimal choice for our model.

% \begin{table}[t]
%     \centering
%     \caption{Choosing Temporal Localization Parameter $t$}
%     \label{tab:experiment-results}
%     \begin{tabular}{cccccc}
%         \toprule
%         \textbf{$t$ Value} & 1 & 2 & 4 & 8 & 16 \\
%         \midrule
%         \textbf{Test AUC} & 0.9257 & \textbf{0.9327} & 0.9221 & 0.9046 & 0.8874 \\
%         \bottomrule
%     \end{tabular}
% \end{table}

\subsection{Ablation Study}

As shown in Table \ref{tab:ablation}, the full DualAttWaveNet achieves the best performance with 83.51\% accuracy, 83.51\% F1 score, and 0.9327 AUC. Removing the mutual attention mechanism ("w/o Mutual Attention") causes consistent performance drops (0.6\% accuracy, 0.6\% F1, 0.3\% AUC). Disabling the wavelet loss ("w/o Wavelet Loss") further degrades accuracy by 1.0\%, proving the necessity of joint time-frequency optimization. The vanilla implementation exhibits significant performance limitations (79.95\% accuracy, 0.9175 AUC), demonstrating a 3.6\% accuracy gap compared to the full model, which highlights the collective contribution of our dual-attention design and spectral-aware constraints to robust classification capability.

\begin{table}[t]
    \caption{Ablation Study of DualAttWaveNet Components}
    \label{tab:ablation}
    \centering

    \begin{tabular}{lccc}
        \toprule
        \textbf{Model Variant} & \textbf{Accuracy (\%)} $\uparrow$ & \textbf{F1 Score} $\uparrow$ & \textbf{AUC} $\uparrow$ \\
        \midrule
        DualAttWaveNet (Full)  & 0.8351                            & 0.8351                       & 0.9327                  \\
        \cmidrule{1-4}
        w/o Mutual Attention   & 0.8289                            & 0.8288                       & 0.9294                  \\
        w/o Wavelet Loss       & 0.8273                            & 0.8273                       & 0.9283                  \\
        Vanilla Implementation & 0.7995                            & 0.7975                       & 0.9175                  \\
        \bottomrule
    \end{tabular}

\end{table}

\section{Conclusion}
\label{sec:conclusion}

In this paper, we present DualAttWaveNet, a computationally-efficient multimodal framework for interference detection in GSO/LEO coexistence systems. We used a bidirectional attention mechanism to fuse time and frequency domain signals. This design integrates the capability of attention module and at the same time uses less computational resources. Additionally, we introduced a wavelet loss besides traditional MSE loss to enforce the reconstructed output to be consistent across all scales. Extensive evaluations in synthetic Ku-band scenarios show DualAttWaveNet achieves 0.9327 AUC with 83.51\% accuracy, surpassing state-of-the-art baselines TrID. The model is trained and tested on an edge NVIDIA 3050Ti GPU, demonstrating 50\% faster inference time compared to TrID while maintaining competitive F1 score. Future work will optimize spectral downsampling strategies for different spectrum overlaps scenarios.

\bibliographystyle{IEEEtran}
\bibliography{references}

\end{document}